# Superconducting nanowire single photon detection system for space applications

**Lixing You,**[1,2,5] **Jia Quan,**[3] **Yong Wang,**[1,4] **Yuexue Ma,**[3] **Xiaoyan Yang**[1,2]**, Yanjie Liu,**[3] **Hao Li,**[1,2] **Jianguo Li,**[3] **Juan Wang,**[3] **Jingtao Liang,**[3,6] **Zhen Wang**[1,2] **and Xiaoming Xie**[1,2]

[1]*State Key Laboratory of Functional Materials for Informatics, Shanghai Institute of Microsystem and Information Technology (SIMIT), Chinese Academy of Sciences (CAS), Shanghai 200050, China*
[2]*Center for ExcelleNce in Superconducting Electronics (CENSE), Chinese Academy of Sciences (CAS), Shanghai 200050, China*
[3]*Key Laboratory of Space Energy Conversion Technologies, Technical Institute of Physics and Chemistry (TIPC), Chinese Academy of Sciences (CAS), Beijing 100190, China*
[4]*University of Chinese Academy of Sciences, Beijing 100049, China*
[5]*lxyou@mail.sim.ac.cn*
[6]*jtliang@mail.ipc.ac.cn*

**Abstract:** Superconducting nanowire single photon detectors (SNSPDs) have advanced various frontier scientific and technological fields such as quantum key distribution and deep space communications. However, limited by available cooling technology, all past experimental demonstrations have had ground-based applications. In this work we demonstrate a SNSPD system using a hybrid cryocooler compatible with space applications. With a minimum operational temperature of 2.8 K, this SNSPD system presents a maximum system detection efficiency of over 50% and a timing jitter of 48 ps, which paves the way for various space applications.

**References and links**

1. F. Marsili, V. B. Verma, J. A. Stern, S. Harrington, A. E. Lita, T. Gerrits, I. Vayshenker, B. Baek, M. D. Shaw, R. P. Mirin, and S. W. Nam, "Detecting single infrared photons with 93% system efficiency," Nat. Photon. **7**, 210-214 (2013).
2. W. Zhang, L. You, H. Li, J. Huang, C. Lv, L. Zhang, X. Liu, J. Wu, Z. Wang, and X. Xie, "NbN superconducting nanowire single photon detector with efficiency over 90% at 1550 nm wavelength operational at compact cryocooler temperature," Sci. China-Phys. Mech. Astron. **60**, 120314 (2017).
3. X. Y. Yang, H. Li, W. J. Zhang, L. X. You, L. Zhang, X. Y. Liu, Z. Wang, W. Peng, X. M. Xie, and M. H. Jiang, "Superconducting nanowire single photon detector with on-chip bandpass filter," Opt. Express **22**, 16267-16272 (2014).
4. N. Calandri, Q.-Y. Zhao, D. Zhu, A. Dane, and K. K. Berggren, "Superconducting nanowire detector jitter limited by detector geometry," Appl. Phys. Lett. **109**, 152601 (2016).
5. J. Wu, L. You, S. Chen, H. Li, Y. He, C. Lv, Z. Wang, and X. Xie, "Improving the timing jitter of a superconducting nanowire single-photon detection system," Appl. Opt. **56**, 2195-2200 (2017).
6. H.-L. Yin, T.-Y. Chen, Z.-W. Yu, H. Liu, L.-X. You, Y.-H. Zhou, S.-J. Chen, Y. Mao, M.-Q. Huang, W.-J. Zhang, H. Chen, M. J. Li, D. Nolan, F. Zhou, X. Jiang, Z. Wang, Q. Zhang, X.-B. Wang, and J.-W. Pan, "Measurement-Device-Independent Quantum Key Distribution Over a 404 km Optical Fiber," Phys. Rev. Lett. **117**, 190501 (2016).
7. H. Li, S. Chen, L. You, W. Meng, Z. Wu, Z. Zhang, K. Tang, L. Zhang, W. Zhang, X. Yang, X. Liu, Z. Wang, and X. Xie, "Superconducting nanowire single photon detector at 532 nm and demonstration in satellite laser ranging," Opt. Express **24**, 3535-3542 (2016).
8. L. Xue, Z. Li, L. Zhang, D. Zhai, Y. Li, S. Zhang, M. Li, L. Kang, J. Chen, P. Wu, and Y. Xiong, "Satellite laser ranging using superconducting nanowire single-photon detectors at 1064 nm wavelength," Opt. Lett. **41**, 3848-3851 (2016).
9. http://www.shicryogenics.com/products/4k-cryocoolers/rdk-101d-4k-cryocooler-series/
10. http://photonspot.com/cryogenics/sorption-fridge
11. http://www.scontel.ru/sspd/
12. D. V. Murphy, J. E. Kansky, M. E. Grein, R. T. Schulein, M. M. Willis, and R. E. Lafon, "LLCD operations using the Lunar Lasercom Ground Terminal," Proc. SPIE 8971, 89710V (2014).
13. V. Kotsubo, R. Radebaugh, P. Hendershott, M. Bonczyski, B. Wilson, S. W. Nam, and J. N. Ullom, "Compact 2.2 K Cooling System for Superconducting Nanowire Single Photon Detectors," IEEE T Appl. Supercond. **27**, 1-5 (2017).
14. X. Yang, L. You, L. Zhang, C. Lv, H. Li, X. Liu, H. Zhou, and Z. Wang, "Comparison of superconducting nanowire single photon detectors made of NbTiN and NbN thin films," IEEE T Appl. Supercond. 10.1109/TASC.2017.2776288 (2017 in press).
15. H. Li et al, State Key Laboratory of Functional Materials for Informatics, Shanghai Institute of Microsystem and Information Technology, Chinese Academy of Sciences, Shanghai 200050, China, are preparing a manuscript to be called "Improving detection efficiency of superconducting nanowire single photon detector with multi-layer anti-reflection coating."
16. R. F. Boyle and R. G. Ross "Overview of NASA space cryocooler programs" in *Advances in Cryogenic Engineering: Proceedings of the Cryogenic Engineering Conference*, **47**,1037-1044 (2002).
17. J. H. Cai, M. G. Zhao, and G. T. Hong, "The Pulse Tube Cryocooler of GF-4 Satellite Staring Camera", Spacecraft Recovery & Remote Sensing **37**(4), 66-71 (2016) (in Chinese)
18. L. You, X. Yang, Y. He, W. Zhang, D. Liu, W. Zhang, L. Zhang, L. Zhang, X. Liu, S. Chen, Z. Wang, and X. Xie, "Jitter analysis of a superconducting nanowire single photon detector," AIP Adv. **3**, 072135 (2013).
19. F. Najafi, A. Dane, F. Bellei, Q. Zhao, K. A. Sunter, A. N. McCaughan, and K. K. Berggren, "Fabrication Process Yielding Saturated Nanowire Single-Photon Detectors with 24-ps Jitter," IEEE J Sel. Top. Quant. **21**, 1-7 (2015).

## 1. Introduction

Superconducting nanowire single photon detectors (SNSPDs) outperform their semiconducting counterparts with a higher system detection efficiency (*SDE*) [1,2], a low dark count rate (*DCR*) [3], and low timing jitter [4,5], which has been demonstrated in various applications, such as quantum key distribution [6] and satellite laser ranging [7,8]. Due to the need for low operational temperatures, the cryogenic partisan is an indispensable component of a practical SNSPD system. A popular choice is either a compact mechanical cryocooler or a liquid helium dewar. The former can provide an operating temperature as low as 2.1 K, such as the two-stage Gifford–McMahon (G-M) cryocooler (Model No. RDK-101D from Sumitomo Inc.) [9]. Embedded with an extra sorption fridge, the system can reach a temperature as low as 0.8 K for a limited period of time [10]. A liquid helium dewar, together with a pump and capillary, can cool SNSPDs down to 1.7 K [11]. However, both of the above systems can only operate on the ground, which limits the use of SNSPDs to ground-based applications. For example, in the Lunar Laser Communication Demonstration project of the National Aeronautics and Space Administration, G-M cryocooler-based SNSPD systems were adopted at the employed ground station. Meanwhile, semiconducting single photon detectors without complicated cryocoolers were used for the satellite [12].

To extend the use of SNSPDs to satellite- and space-based applications, a compatible cryocooler needs to be developed. Because the performance parameters, such as the *SDE* and *DCR*, of SNSPDs is extremely sensitive to the operational temperature [2], the temperature for such a cryocooler should be lower than 4 K; the lower the temperature, the better the performance. However, reaching a temperature of approximately 2 K in space with a long-life mechanical cooler is still a significant technical challenge [13]. In addition to the need for such low temperatures, the size, weight, and power (SWaP) of the system also needs to be considered.

In this paper, we present a hybrid cryocooler that is compatible with space applications, which incorporates a two-stage high-frequency pulse tube (PT) cryocooler and a $^4$He Joule–Thomson (JT) cooler. To make a practical SNSPD system for space applications, we chose a superconducting NbTiN ultrathin film, which can operate sufficiently well above 2 K, to fabricate the SNSPDs, instead of using WSi, which usually requires sub-1-K temperatures [1]. The hybrid cryocooler successfully cooled an NbTiN SNSPD down to a minimum temperature of 2.8 K. The NbTiN SNSPD showed a maximum *SDE* of over 50% at a wavelength of 1550 nm and a *SDE* of 47% at a *DCR* of 100 Hz. Therefore, these results experimentally demonstrate the feasibility of space applications for this SNSPD system.

## 2. SNSPD fabrication and packaging

The SNSPDs were made of an NbTiN film deposited on a 2-inch double-sided thermal-oxidized Si substrate. Before the NbTiN deposition, an anti-reflection coating with dielectric layers was ion-beam-assisted deposited on the backside of the substrate; this provides better transmission than a conventional $SiO_2$ layer with a thickness of a 1/4 wavelength. The nominal NbTiN film thickness was determined by the sputtering rate and time. Here the target nominal thickness of the NbTiN film was empirically chosen to be 5 nm. Subsequently, a meandered nanowire covering a circular area with a diameter of 15 μm was structured into the NbTiN film using e-beam lithography and reactive ion etching. The linewidth and the period of the nanowire were 70 nm and 160 nm, respectively. To form the top optical cavity to enhance the absorption, a 200-nm-thick SiO layer was deposited on top of the nanowire as a dielectric material and then a 5-nm/100-nm-thick Ti/Au film was deposited on the SiO layer to create a mirror for the cavity. The detailed design fabrication has been described elsewhere [14,15].

To optically couple the SNSPD to the fiber, we adopted a backside optical coupling package. A lensed single-mode fiber was manually aligned with the active area of the SNSPD from the backside of the chip with an alignment error of less than 3 μm. Then, the packaged detector was mounted and characterized inside either a conventional G-M cryocooler or the hybrid JT cryocooler described in Section 3. To measure the SNSPD, a bias-tee and a low noise amplifier (LNA) were adopted at room temperature. An isolated voltage source in series with a resistor (20 kΩ) provided a stable current bias to the detector. The bias current was fed into the device through the *dc* port of the bias-tee and the electric response pulses of the SNSPD were extracted from the *ac* port of the bias-tee and amplified by the LNA. The amplified pulse signals could be monitored via an oscilloscope or counted via a photon counter. The fiber laser source was connected with two variable attenuators and a polarization controller in series. The performance of the SNSPD is described in detail in Section 4.

## 3. Hybrid cryocooler for space applications

The hybrid cryocooler is composed of two thermally coupled parts: a two-stage high-frequency PT cryocooler and a JT loop, as illustrated in Fig. 1. It has no moving parts at low temperature and therefore has a long lifetime and high reliability as required for space applications. The high-frequency PT cryocooler is characterized by its small size, long life, and low vibration, and is therefore widely used in space applications [16]. The linear JT compressors are the inheritance of the PT compressors, which have been successfully used in satellites [17]. The two cold fingers of the high-frequency PT cooler, i.e., the precooling JT loops at 65 K and 15 K, are of a thermally coupled coaxial configuration. The JT loop is composed of a three-stage JT compressor system, counter-flow heater exchangers, precooling heat exchangers, a JT valve, and an evaporator. The high pressure $^4$He gas passes through the inner tube of the three counter-flow heat exchangers where it is cooled by the low pressure returning gas. Two additional heat exchangers are mounted on the cold ends of the precooler to provide precooling power for the JT cycle. When helium gas goes through the JT orifice, a fraction of the helium gas condenses into liquid during the expansion process. The device mounted on the evaporator is cooled by the liquid helium produced via the JT effect. The three-stage JT compressor system operates at a suction pressure much lower than atmospheric pressure, which enables temperatures lower than 3 K to be achieved. Both the PT cryocooler and the JT cycle are driven by oil-free linear compressors, which make the hybrid JT cooler suitable for space applications.

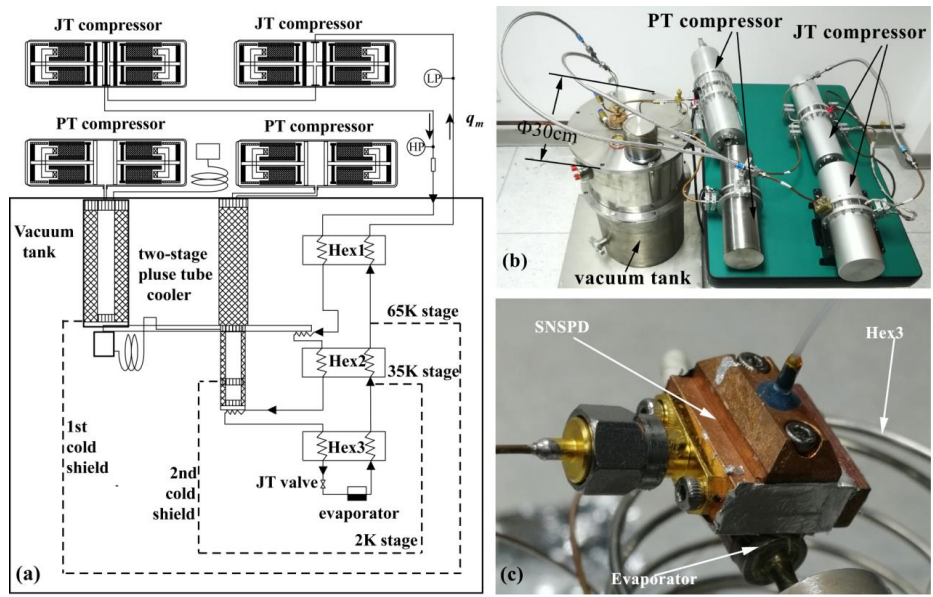

Fig. 1. (a) Schematics of the hybrid JT cryocooler; (b) A photo of the SNSPD system, including the hybrid JT cryocooler; (c) A photo of the SNSPD package installed on the evaporator of the JT cooler.

Figure 2 shows the cooling down curves of the hybrid JT cryocooler. First, the two-stage PT cryocooler is started to precool the JT loop. The first and second stage cold heads of the PT cryoooler reach their operating temperatures in approximately 10 hr and 18 hr, respectively. When the inlet helium gas is under 40 K, the JT compressors are started to generate a pressure difference. Then, the temperatures before and after the JT orifice gradually fall faster. Detailed characteristics of the temperatures before and after the JT are given in the inset in Fig. 2. The operating and performance parameters of the hybrid JT cryocooler are presented in Table 1. When the total power consumption of the hybrid JT cryocooler is 319.8 W (279.2 W for the PT cooler and 40.6 W for the JT cooler), a minimum temperature of 2.80 K is achieved and remains stable throughout the testing process of the SNSPD. Note that during the test at another location without the installation of the SNSPD, a minimum temperature of 2.6 K was obtained. The total weight of the cryocooler system is 55 kg.

**Table 1. Operating and performance parameters of the hybrid JT cryocooler**

| Parameters of the hybrid cryocooler | | First stage | Second stage | Third stage |
|---|---|---|---|---|
| Pulse tube cryocooler | Input electrical power (W) | 168.8 | 110.4 | - |
| | Frequency (Hz) | 42 | 18 | - |
| | Precooling temperature (K) | 65.7 | 15.3 | - |
| JT cooler | Input electrical power (W) | 9.5 | 10.2 | 20.9 |
| | Frequency (Hz) | 50 | 50 | 50 |
| | Pressure ratio $P_h/P_l$ (MPa/MPa) | | 0.6/0.018 | |
| Temperature of the JT cold end (K) | | | 2.80 | |

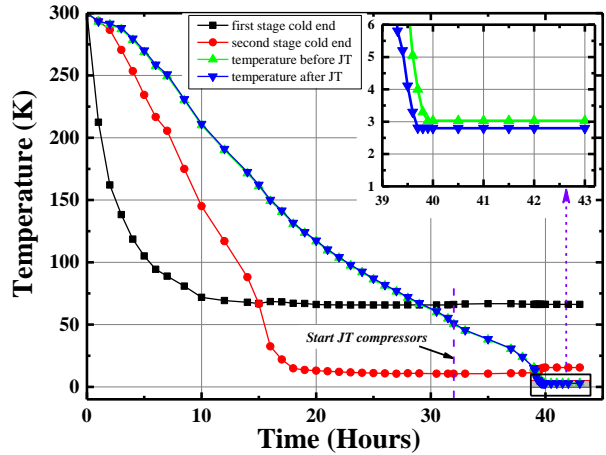

Fig. 2. Cooling down temperature curves of the hybrid JT cryocooler. The inset shows the enlarged curves from hour 39 to hour 43. The evaporator reaches a minimum temperature of 2.8 K.

## 4. Performance of the SNSPD

Because the lowest temperature of the hybrid cryocooler (2.8 K) is higher than that of a conventional G-M cryocooler (2.1 K), performance differences of SNSPDs cooled with the two different cryocoolers are expected. The SNSPD was first measured using a conventional

G-M cryocooler for reference and then measured using the hybrid cryocooler. The characteristics of the SNSPDs were measured and are discussed in comparison.

The detector has a superconducting transition temperature of 7.9 K. Fig. 3(a) shows the current–voltage (*I*–*V*) curves of the SNSPDs. Both *I*–*V* curves have similar behaviors, but different switching currents ($I_{sw}$: 19.6 µA and 22.0 µA, respectively) due to the different operating temperatures (2.8 K and 2.1 K, respectively). Fig. 3(b) shows the bias current dependence of the *SDE* and *DCR*. The SNSPD cooled with the hybrid cryocooler shows a maximum *SDE* of over 50%. The *SDE* was 47% when *DCR* = 100 Hz. For comparison, the SNSPD in the G-M cryocooler shows an *SDE* of ~90% when *DCR* = 100 Hz. On the other hand, with the same bias current, the absolute difference in the *SDE* values measured using the two cryocoolers is less than 10%, which is useful to estimate the maximum *SDE* at different temperatures when the bias current is known. Fig. 3 (c) shows the instant waveforms of the pulses when the bias current is $0.93I_{sw}$ (*DCR* = 100 Hz). Similar shapes and noise levels, but different amplitudes of the response pulses, were observed due to the different switching currents. The different amplitudes caused different signal to noise ratios (SNRs), which resulted in different timing jitter values [18], as shown in Fig. 3(d). The SNSPD cooled with the hybrid cryocooler has a full-width half maximum timing jitter of 48 ps, 20% larger than the value (40 ps) of the SNSPD cooled with the G-M cryocooler.

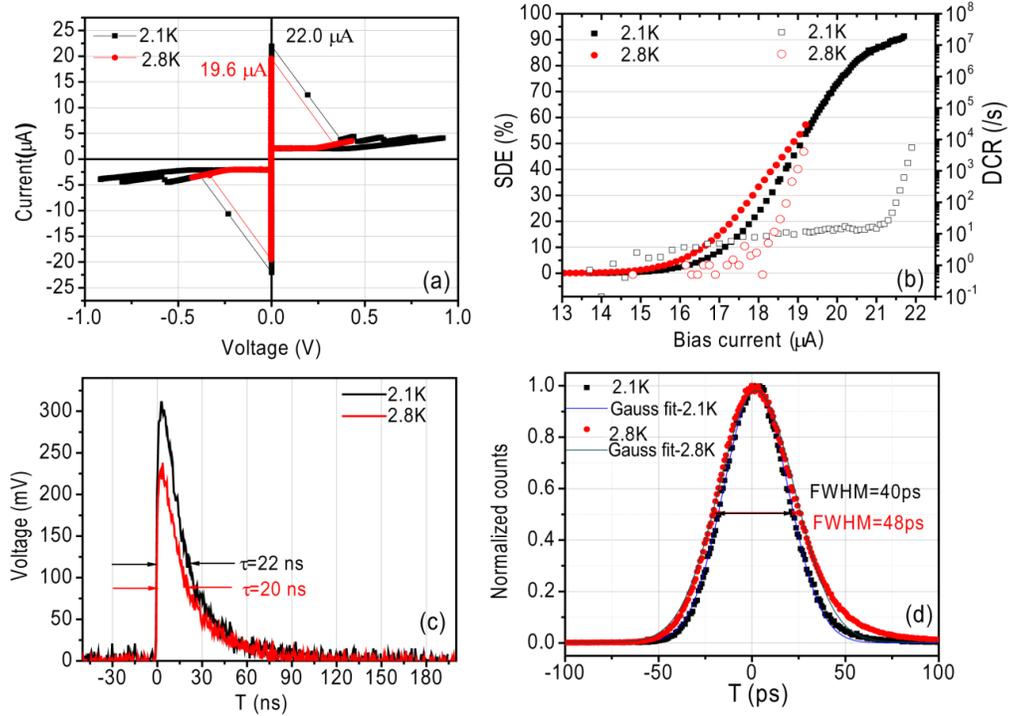

Fig. 3. Performance comparison of the SNSPD cooled with the hybrid cryocooler and the G-M cryocooler: (a) *I*–*V* curves; (b) *SDE* and *DCR* versus the bias current $I_b$; (c) the transient amplified response pulses (the amplification ratio is 300); and (d) the measured timing jitters with Gauss-fit curves when $I_b = 0.93I_{sw}$.

## 5. Discussion

We have demonstrated the operation of an SNSPD cooled with a hybrid cryocooler that is compatible with space applications. In addition, we believe that the performance of the system can be further improved. First, the optimization of the JT compressors and the PT coolers can decrease the temperature to less than 2.5 K, which will further improve the performance of the SNSPD. Furthermore, temperatures lower than 2 K can be readily achieved by replacing the working medium of the JT loop with $^3$He for space applications, where the cost is not the chief limiting factor. However, there is still room for improvement considering the SWaP of the present cryocooler. Second, previous studies on SNSPDs have focused on improving the performance at 2.1 K and even lower temperatures. It would be interesting to optimize the performance of the SNSPD at a higher temperature that the hybrid cryocooler can reach, for example, 2.5 K, 2.8 K, or even higher. Therefore, we may be able to relax this stringent requirement for the cryocooler. In addition, advancing technologies in the detector's design and fabrication may further improve its performance. For example, we may increase the SNR, then decrease the timing jitter by parallelizing the nanowires [19].

## 6. Conclusions

We have successfully demonstrated a NbTiN SNSPD cooled with a hybrid cryocooler that is compatible with space applications. The hybrid cryocooler reached a minimum temperature of 2.8 K using 319.8 W of electrical power, while the total weight was 55 kg. The key technologies used in the precooling PT cryocooler and the JT compressor were flight-proven [17]. Even though there is still significant room for improving the SWaP of the hybrid cryocooler, the general SWaP of the present prototype is within the range of acceptance as a payload for a satellite. At the minimum temperature of 2.8 K, the SNSPD system had a maximum *SDE* of over 50% and an *SDE* of 47% at a *DCR* of 100 Hz, while the timing jitter was 48 ps. As a next step, space environment tests and consequent improvements to the SNSPDs will be conducted. The results obtained will pave the way for future space application of SNSPD, such as space quantum key distribution, deep space communication, and light detection and ranging.

## Funding

National Key R&D Program of China (2017YFA0304000); National Natural Science Foundation of China (61671438, 61401443, 61501439, 51776213); Science and Technology Commission of Shanghai Municipality (16JC1400402); National Basic Research Program of China (613322).